\documentclass[preprint,superscriptaddress,amsmath,amssymb,prd,aps,showpacs,floatfix,nofootinbib]{revtex4-1}
\usepackage{graphicx}
\usepackage{dcolumn}
\usepackage{bm}
\usepackage{mathrsfs}
\usepackage{hyperref}
\usepackage{amsmath}
\usepackage{amssymb}
\usepackage{float}
\usepackage{dcolumn}
\usepackage{multirow}

\begin{document}
\title{Description of the $\Xi_c$ and $\Xi_b$ states as molecular states}
\date{\today}
\author{Q.~X.~Yu}
\email{qixinyu@ific.uv.es}\email{yuqx@mail.bnu.edu.cn}
\affiliation{College of Nuclear Science and Technology, Beijing Normal University, Beijing 100875, China}
\affiliation{Departamento de
F\'{\i}sica Te\'orica and IFIC, Centro Mixto Universidad de
Valencia-CSIC Institutos de Investigaci\'on de Paterna, Aptdo.22085,
46071 Valencia, Spain}

\author{R.~Pavao}
\email{rpavao@ific.uv.es}
\affiliation{Departamento de
F\'{\i}sica Te\'orica and IFIC, Centro Mixto Universidad de
Valencia-CSIC Institutos de Investigaci\'on de Paterna, Aptdo.22085,
46071 Valencia, Spain}

\author{V.~R.~Debastiani}
\email{vinicius.rodrigues@ific.uv.es}
\affiliation{Departamento de
F\'{\i}sica Te\'orica and IFIC, Centro Mixto Universidad de
Valencia-CSIC Institutos de Investigaci\'on de Paterna, Aptdo.22085,
46071 Valencia, Spain}

\author{E.~Oset}
\email{oset@ific.uv.es}
\affiliation{Departamento de
F\'{\i}sica Te\'orica and IFIC, Centro Mixto Universidad de
Valencia-CSIC Institutos de Investigaci\'on de Paterna, Aptdo.22085,
46071 Valencia, Spain}

\begin{abstract}
In this work we study several $\Xi_c$ and $\Xi_b$ states dynamically generated from the meson-baryon interaction in coupled channels, using an extension of the local hidden gauge approach in the Bethe-Salpeter equation. These molecular states appear as poles of the scattering amplitudes, and several of them can be identified with the experimentally observed $\Xi_c$ states, including the $\Xi_c(2790)$, $\Xi_c(2930)$, $\Xi_c(2970)$, $\Xi_c(3055)$ and $\Xi_c(3080)$. Also, for the recently reported $\Xi_b(6227)$ state, we find two poles with masses and widths remarkably close to the experimental data, for both the $J^P=1/2^-$ and $J^P=3/2^-$ sectors.
\end{abstract}

\maketitle


\section{Introduction}

Heavy baryons containing one $c$ or $b$ quark have been the subject of intense study. Starting from early quark models \cite{Capstick:1986bm}, work along this line has been rather extensive and fruitful \cite{Ebert:2007nw,Garcilazo:2007eh,Ebert:2011kk,Ortega:2012cx,Shah:2016nxi,Santopinto:2018ljf,Valcarce:2008dr}. QCD lattice has also contributed to this area \cite{Bowler:1996ws,Burch:2008qx,Brown:2014ena} and dynamical models building molecular states in coupled meson-baryon channels \cite{GarciaRecio:2012db,Liang:2014eba,Liang:2014kra,Romanets:2012hm,Liu:2018bkx,Huang:2018wgr,Chen:2017xat,Dong:2010xv,Dong:2014ksa,Ortega:2014eoa,He:2015cea,Chen:2017vai,Liu:2018zzu} have also brought their share to this intense research. There are also many review papers on the subject to which we refer the reader \cite{Korner:1994nh,Bianco:2003vb,Klempt:2009pi,Crede:2013sze,Cheng:2015iom,Chen:2016qju,Chen:2016spr,Guo:2017jvc}.

In the present work we study in detail the $\Xi_c$ and $\Xi_b$ states from the molecular point of view. There are many  $\Xi_c$ states reported in the PDG \cite{Tanabashi:2018oca} corresponding to excited states. One of the $\Xi_c$ states, $\Xi_c(2930)$, first reported by the BaBar Collaboration \cite{Aubert:2007eb}, was recently confirmed with more statistics by the Belle Collaboration \cite{Li:2017uvv}. On the other hand, for $\Xi_b$, apart from the $J^P=1/2^+$ ground states, $\Xi_b$, $\Xi^\prime_b$, and the $J^P=3/2^+$ $\Xi_b^*$, there are no states reported in the PDG \cite{Tanabashi:2018oca}. Yet, the LHCb Collaboration has recently reported one such state, the $\Xi_b(6227)$ \cite{Aaij:2018yqz}, which we shall also investigate in the present work.

Recent studies of such states using QCD sum rules can be found in Refs.~\cite{Chen:2017sci,Chen:2016phw,Azizi:2018dva,Wang:2018alb}, where also reference to works on this particular issue is done, mostly on quark models. As to molecular states of this type we refer to the work of Ref.~\cite{Romanets:2012hm}.

The experimental finding of five new excited $\Omega_c$ states by the LHCb Collaboration \cite{Aaij:2017nav} (see also Ref.~\cite{Yelton:2017qxg}) stimulated new work along the molecular line and in Ref.~\cite{Montana:2017kjw} a study was done of coupled channels interaction using and extension to $SU(4)$ of the chiral Lagrangians. The interesting result from this work was that two states could be interpreted as $1/2^-$ resonances and the mass and width were well reproduced. This is a non trivial achievement since in other approaches mostly masses are studied and not widths. Some quark models go one step forward and using the $^3P_0$ model also evaluate widths, as in Ref.~\cite{Santopinto:2018ljf}. The fact that the widths obtained are quite different than in the molecular model is a positive sign that the study of the widths, and partial decay widths of these resonances, carry valuable information concerning their nature.

The work of Ref.~\cite{Montana:2017kjw}, with vector meson exchange in an extension to $SU(4)$ of the chiral Lagrangians, got a boost from Ref.~\cite{Debastiani:2017ewu}, where it was shown that the relevant matrix elements of the interaction can be obtained considering the exchange of light vector mesons in an extension of the local hidden gauge approach \cite{Bando:1984ej,Bando:1987br,Meissner:1987ge,Harada:2003jx,Nagahiro:2008cv}, where the heavy quarks were mere spectators, such that there was no need to invoke $SU(4)$ and one could make a mapping of the $SU(3)$ results where the local hidden gauge approach was developed. Like in Ref.~\cite{Montana:2017kjw}, in Ref.~\cite{Debastiani:2017ewu} the same two states were obtained with similar widths, and in addition there was another state reproduced with $3/2^-$, which was not addressed in \cite{Montana:2017kjw}. Similar results were then obtained in Ref.~\cite{Nieves:2017jjx} with a continuation of the work of Ref.~\cite{Romanets:2012hm} with parameters adjusted to input from the experiment of Ref.~\cite{Aaij:2017nav}.

The former results stimulated further work along these lines with predictions for $\Omega_b$ states in Ref.~\cite{Liang:2017ejq}, for which there are not yet experimental counterparts. Encouraged by the success in the $\Omega_c$ states, in the present work we follow this line of research to study $\Xi_c$ and $\Xi_b$ states. In the first case there are several states to compare with our predictions, and in the second case only one excited state, such that many of the states found will be predictions to be tested with future experiments.

Related to these works is the study of the $\Xi_{cc}$ molecular states in Ref.~\cite{Dias:2018qhp}, stimulated by the new measurement of the $\Xi_{cc}$ by the LHCb Collaboration, with a mass of $3621\,\rm MeV$ \cite{Aaij:2017ueg}.
This value is higher than that previously measured by the SELEX Collaboration \cite{selex1,selex2}. However, this first measurement by SELEX was not confirmed by the FOCUS \cite{Ratti:2003ez}, Belle \cite{Aubert:2006qw}, BABAR \cite{Chistov:2006zj} and the LHCb \cite{Aaij:2013voa} Collaborations. Using the value of the new measurement of the LHCb Collaboration \cite{Aaij:2017ueg}, molecular $\Xi_{cc}$ states were studied in Ref.~\cite{Dias:2018qhp}, where excited bound states were found above $4000\,\rm MeV$ and broad $\Xi_{cc} \pi$ and $\Xi^\ast_{cc} \pi$ resonances were found around $3837$ and $3918$ MeV, respectively.

With all this recent experimental activity there is much motivation to make predictions with different models which can serve as potential guide for  experimental set ups and finally to deepen our understanding of the nature of the baryon resonances.

\section{Formalism}

Recently two new resonances, the $\Xi_c(2930)$ $\left(J=?^?\right)$ and the $\Xi_b(6227)$ $\left(J=?^? \right)$, have been measured by the Belle~\cite{Li:2017uvv} and LHCb~\cite{Aaij:2018yqz} Collaborations, respectively. Besides these, there is also an abundance of other unexplained resonances in the charm sector~\cite{Tanabashi:2018oca}: $\Xi_c(2790)$ $\left(J=1/2^-\right)$, $\Xi_c(2815)$ $\left(J=3/2^-\right)$, $\Xi_c(2970)$ $\left(J=?^?\right)$, $\Xi_c(3055)$ $\left(J=?^?\right)$, $\Xi_c(3080)$ $\left(J=?^?\right)$ and $\Xi_c(3123)$ $\left(J=?^?\right)$. The objective of this work is to shed some light into the nature of these states and to explain at least some of them within the hadronic molecular picture. For that, we shall use an extension of the chiral unitary approach with coupled channels outlined in Ref.~\cite{Debastiani:2017ewu}, since as we shall see, only the light quarks play a relevant role in the interaction. As in Ref.~\cite{Debastiani:2017ewu}, we will separate the interaction into pseudoscalar meson-baryon$(1/2^+)$ ($PB$), vector meson-baryon$(1/2^+)$ ($VB$) and pseudoscalar meson-baryon$(3/2^+)$ ($PB^*$). One should mention that in this theory, these three sectors do not decay into each other, because that would require the exchange of pseudoscalar mesons, and those transitions are momentum-dependent and small compared to the ones with a vector meson exchange \cite{Debastiani:2017ewu,Xiao:2013yca}. Analyzing the spin-parity of each sector, we find that for the states that arise from $PB$ we have $J^P = 0^- \otimes 1/2^+ = 1/2^-$, for $VB$ we have degenerate states $J^P =1^- \otimes 1/2^+ = 1/2^-, \ 3/2^- $ and for $PB^*$ we have $J^P =0^- \otimes 3/2^+ = 3/2^- $. In Tables~\ref{tab:thPBc} to~\ref{tab:thPBsc} we show, for the charm sector, the channels chosen for the $PB$, $VB$ and $PB^*$ sectors. To get the channels for the beauty sector, one needs only to substitute the $c$ quark by a $b$ quark, and we show the results in tables~\ref{tab:thPBb} to~\ref{tab:thPBsb}.

\begin{table*}[h!]
\renewcommand\arraystretch{1.2}
\centering
\caption{\vadjust{\vspace{-2pt}}Charm sector channels with $J^P=1/2^-$ and respective thresholds.}\label{tab:thPBc}
\begin{tabular*}{1.00\textwidth}{@{\extracolsep{\fill}}cccccccccccc}
\hline
\hline
\textbf{Channel}& $\Xi_c\pi$ & $\Xi_c^\prime\pi$ & $\Lambda_c\bar K$ & $\Sigma_c\bar K$ & $\Lambda D$ & $\Xi_c\eta$ & $\Sigma D$ & $\Xi_c^\prime\eta$ & $\Omega_c K$ & $\Xi D_s$ \\
\hline
\textbf{Threshold (MeV)}& 2607 & 2716 & 2782 & 2949 & 2983 & 3017 & 3060 & 3126 & 3191 & 3287\\
\hline
\hline
\end{tabular*}
\end{table*}
\begin{table*}[h!]
\renewcommand\arraystretch{1.2}
\centering
\caption{\vadjust{\vspace{-2pt}}Charm sector channels with $J^P=1/2^-, \ 3/2^-$ and respective thresholds.}\label{tab:thVBc}
\begin{tabular*}{1.00\textwidth}{@{\extracolsep{\fill}}cccccccc}
\hline
\hline
\textbf{Channel}& $\Lambda D^*$ & $\Lambda_c\bar K^*$ & $\Sigma D^*$ & $\Xi_c\rho$ & $\Xi_c\omega$ & $\Sigma_c\bar K^*$ \\
\hline
\textbf{Threshold (MeV)}& 3124 & 3182 & 3202 & 3245 & 3252 & 3349\\
\hline
\hline
\end{tabular*}
\end{table*}
\begin{table*}[h!]
\renewcommand\arraystretch{1.2}
\centering
\caption{\vadjust{\vspace{-2pt}}Charm sector channels with $J^P=3/2^-$ and respective thresholds.}\label{tab:thPBsc}
\begin{tabular*}{1.00\textwidth}{@{\extracolsep{\fill}}ccccccc}
\hline
\hline
\textbf{Channel} & $\Xi_c^*\pi$ & $\Sigma_c^*\bar K$ & $\Xi_c^*\eta$ & $\Sigma^*D$ & $\Omega_c^*K$ \\
\hline
\textbf{Threshold (MeV)} & 2784 & 3014 & 3194 & 3252 & 3262 \\
\hline
\hline
\end{tabular*}
\end{table*}
\begin{table*}[h!]
\renewcommand\arraystretch{1.2}
\centering
\caption{\vadjust{\vspace{-2pt}}Beauty sector channels with $J^P=1/2^-$ and respective thresholds.}\label{tab:thPBb}
\begin{tabular*}{1.00\textwidth}{@{\extracolsep{\fill}}cccccccccccc}
\hline
\hline
\textbf{Channel} & $\Xi_b\pi$ & $\Xi_b^\prime\pi$ & $\Lambda_b\bar K$ & $\Sigma_b\bar K$ & $\Lambda \Bar B$ & $\Xi_b\eta$ & $\Sigma \bar B$ & $\Xi_b^\prime\eta$ & $\Omega_b K$ & $\Xi B_s$ \\
\hline
\textbf{Threshold (MeV)}& 5931 & 6073 & 6115 & 6309 & 6395 & 6341 & 6473 & 6483 & 6542 & 6685\\
\hline
\hline
\end{tabular*}
\end{table*}
\begin{table*}[h!]
\renewcommand\arraystretch{1.2}
\centering
\caption{\vadjust{\vspace{-2pt}}Beauty sector channels with $J^P=1/2^-, \ 3/2^-$ and respective thresholds.}\label{tab:thVBb}
\begin{tabular*}{1.00\textwidth}{@{\extracolsep{\fill}}cccccccc}
\hline
\hline
\textbf{Channel}& $\Lambda \bar B^*$ & $\Lambda_b\bar K^*$ & $\Sigma \bar B^*$ & $\Xi_b\rho$ & $\Xi_b\omega$ & $\Sigma_b\bar K^*$ \\
\hline
\textbf{Threshold (MeV)}& 6440 & 6515 & 6518 & 6568 & 6576 & 6709\\
\hline
\hline
\end{tabular*}
\end{table*}
\begin{table*}[h!]
\renewcommand\arraystretch{1.2}
\centering
\caption{\vadjust{\vspace{-2pt}}Beauty sector channels with $J^P=3/2^-$ and respective thresholds.}\label{tab:thPBsb}
\begin{tabular*}{1.00\textwidth}{@{\extracolsep{\fill}}ccccccc}
\hline
\hline
\textbf{Channel} & $\Xi_b^*\pi$ & $\Sigma_b^*\bar K$ & $\Xi_b^*\eta$ & $\Sigma^*\bar B$ & $\Omega_b^*K$ \\
\hline
\textbf{Threshold(MeV)} & 6091 & 6329 & 6500 & 6664 & 6567 \\
\hline
\hline
\end{tabular*}
\end{table*}

\begin{figure}[tbh]
  \centering
  \includegraphics[width=0.45\textwidth]{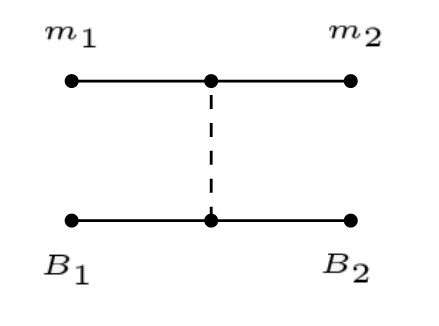}
  \caption{Example of a type of diagram that arises from the LHG.}
  \label{fig:feyn1}
\end{figure}

In this work, the kernel will be calculated using an extension of the local hidden gauge approach (LHG)~\cite{Bando:1984ej,Bando:1987br,Meissner:1987ge,Harada:2003jx,Nagahiro:2008cv}, which produces Feynman diagrams of the type shown in Fig.~\ref{fig:feyn1}, that is, the initial meson baryon pair goes into the final pair through the exchange of a vector meson in the $t-$channel.

In the $PB$ case, the meson-meson interaction (the upper vertex in Fig.~\ref{fig:feyn1}) is given by the $VPP$ Lagrangian:
\begin{equation}
\label{eq:vpp}
\mathcal{L}_{VPP} = -i g \left < \left[ \phi, \partial_{\mu}  \phi \right] V^{\mu} \right>,
\end{equation}
where $g=\displaystyle\frac{m_V}{2 f_{\pi}}$, $\phi$ and $V^{\mu}$ are the $SU(4)$ pseudoscalar meson and vector meson flavor matrices, respectively, and $\left< \cdots  \right>$ is the trace over the $SU(4)$ matrices. Note that the original $\mathcal{L}_{VPP}$ interaction obeys $SU(3)$ flavor symmetry, but just like in Ref.~\cite{Debastiani:2017ewu}, we extend it to $SU(4)$ to take into account the $c$ ($b$) quark. The meson matrices are
\begin{equation}
\phi = \begin{pmatrix}
\frac{1}{\sqrt{2}}\pi^0 + \frac{1}{\sqrt{3}} \eta + \frac{1}{\sqrt{6}}\eta' & \pi^+ & K^+ & \bar{D}^0 \\
 \pi^- & -\frac{1}{\sqrt{2}}\pi^0 + \frac{1}{\sqrt{3}} \eta + \frac{1}{\sqrt{6}}\eta' & K^0 & D^- \\
 K^- & \bar{K}^0 & -\frac{1}{\sqrt{3}} \eta + \sqrt{\frac{2}{3}}\eta' & D_s^- \\
D^0  & D^+ & D_s^+ & \eta_c
\end{pmatrix},
\end{equation}
and
\begin{equation}
\label{eq:matV}
V = \begin{pmatrix}
 \frac{1}{\sqrt{2}}\rho^0 + \frac{1}{\sqrt{2}} \omega & \rho^+ & K^{* +} & \bar{D}^{* 0} \\
 \rho^- & -\frac{1}{\sqrt{2}}\rho^0 + \frac{1}{\sqrt{2}} \omega  & K^{* 0} & \bar{D}^{* -} \\
 K^{* -} & \bar{K}^{* 0}  & \phi & D_s^{* -} \\
 D^{* 0} & D^{* +} & D_s^{* +} & J/\psi
\end{pmatrix}.
\end{equation}

The use of $SU(4)$ in Eq.~\eqref{eq:vpp} is a formality. We shall see later that the dominant terms are due to the exchange of light vectors, where the heavy quark are spectators. Then Eq.~\eqref{eq:vpp} automatically projects over $SU(3)$. The terms with the exchange of a heavy vector are very suppressed, as we shall see. In principle, in this case one would be using explicitly $SU(4)$, however, as seen in Ref.~\cite{Sakai:2017avl}, since the matrices $\phi$ and $V$ stand for $q\bar q$, Eq.~\eqref{eq:vpp} actually only measures the quark overlap of $\phi$ and $V$ and provides a vector structure, hence, the role of $SU(4)$ is just a trivial counting of the number of quarks. The lower vertex $V^\mu BB$ does not rely on $SU(4)$ either, as we see below.

Now, for the lower vertex in Fig.~\ref{fig:feyn1}, the interaction in $SU(3)$ can be described by the following Lagrangian
\begin{equation}
\label{eq:VBB}
\mathcal{L}_{VBB} =  g \left( \overline{B} \gamma_{\mu} \left[V^{\mu} ,B \right]+\left< \overline{B} \gamma_{\mu} B\right> \left< V^{\mu} \right> \right),
\end{equation}
where $B$ is the $SU(3)$ baryon matrix, and $V$ the $3\times 3$ part of $V$ in Eq.~\eqref{eq:matV} containing $\rho$, $\omega$, $K^*$, $\phi$. Here we do a non-relativistic approximation, which consists in substituting $\gamma^{\mu} \rightarrow \gamma^{0}$. The extension to the charm or bottom sectors is done without relying on $SU(4)$ as explained below. As discussed in Refs.~\cite{Debastiani:2017ewu,Dias:2018qhp}, it can be shown that the same interaction in $SU(3)$ of Eq.~\eqref{eq:VBB}can be obtained considering an operator at the quark level, such that Eq.~\eqref{eq:VBB} becomes
\begin{equation}
\label{eq:VBBql}
\mathcal{L}_{VB_fB_i} = g \big< B_f \big|V_{ql} \big| B_i \big>,
\end{equation}
where $ \big| B_i \big>, \  \big| B_f \big>$ are the initial and final baryon spin-flavor wave functions with the following structure,
\begin{equation}\label{eq:BigB}
\big| B\big> = \big| \phi_{\text{flavor}} \big>  \otimes \big| \chi_{\text{spin}} \big>,
\end{equation}
and $V_{ql}$ is the quark operator of the exchanged vector meson, which, for example, for the diagonal ones in Eq.~\eqref{eq:matV} is
\begin{subequations}
\begin{align}
& \rho^0 = \frac{1}{\sqrt{2}} (u \bar{u} - d \bar{d}),\\
& \omega = \frac{1}{\sqrt{2}} (u \bar{u} + d \bar{d}),\\
& \phi = s \bar{s}.
\end{align}
\end{subequations}

The states described by Eq.~\eqref{eq:BigB} are constructed using only $SU(3)$ symmetry, taking the heavy quark as a spectator, which implies that all diagonal terms are described through the exchange of light vectors, respecting heavy quark spin symmetry  (HQSS) \cite{manohar}.

The baryon quark states that we will be using in the $PB$ sector are constructed using the method outlined in Ref.~\cite{Close}, but with the necessary changes in phases in order to obey the sign notation in Ref.~\cite{Miyahara:2016yyh,Pavao:2017cpt}, which is consistent with the chiral matrices. Doing this we obtain the following states:
\begin{subequations}
\allowdisplaybreaks
\begin{align}
&\big| \Xi_c^+ \big> =  \big| \frac{1}{\sqrt{2}} c(us-su) \big> \big| \chi_{MA}\big>,\\
&\big| \Xi_c^0 \big> =  \big| \frac{1}{\sqrt{2}} c(ds-sd) \big> \big| \chi_{MA}\big>,\\
&\big| \Xi_c^{'+} \big> =  \big| \frac{1}{\sqrt{2}} c(us+su) \big> \big| \chi_{MS}\big>,\\
&\big| \Xi_c^{'0} \big> =  \big| \frac{1}{\sqrt{2}} c(ds+sd) \big> \big| \chi_{MS}\big>,\\
&\big|\Lambda_c^+ \big> =  \big| \frac{1}{\sqrt{2}} c(ud-du) \big> \big| \chi_{MA}\big>,\\
&\big|\Sigma_c^{++} \big> =  \big|cuu \big> \big| \chi_{MS}\big>,\\
&\big|\Sigma_c^+ \big> =  \big| \frac{1}{\sqrt{2}} c(ud+du) \big> \big| \chi_{MS}\big>,\\
&\big|\Sigma_c^0 \big> =  \big|cdd \big> \big| \chi_{MS}\big>,\\
&\big|\Lambda^0 \big> =  \frac{1}{\sqrt{2}}\left( \big| \phi_{MS} \big> \big| \chi_{MS} \big> +\big|  \phi_{MA} \big> \big| \chi_{MA} \big> \right),\nonumber\\
& \ \ \ \ \big| \phi_{MS} \big> = \frac{1}{2} \left( \big| dus \big> + \big| dsu \big> - \big| uds \big> - \big| usd \big> \right),\nonumber\\
& \ \ \ \ \big| \phi_{MA} \big> = \frac{1}{2 \sqrt{3}} \left(\big|u (ds-sd) \big> + \big| d(su-us) \big> - 2 \big| s(ud-du) \big> \right),\\
&\big|\Sigma^+ \big> =  \frac{1}{\sqrt{2}}\left( \big| \phi_{MS} \big> \big| \chi_{MS} \big> +\big|  \phi_{MA} \big> \big| \chi_{MA} \big> \right),\nonumber\\
& \ \ \ \ \big| \phi_{MS} \big> = -\frac{1}{\sqrt{6}} \left( \big| u(us+su) \big> -2 \big| s uu \big> \right),\nonumber\\
& \ \ \ \ \big| \phi_{MA} \big> = \frac{1}{\sqrt{2}}\big|u (su-us) \big>,\\
&\big|\Sigma^0 \big> =  \frac{1}{\sqrt{2}}\left( \big| \phi_{MS} \big> \big| \chi_{MS} \big> +\big|  \phi_{MA} \big> \big| \chi_{MA} \big> \right),\nonumber\\
& \ \ \ \ \big| \phi_{MS} \big> = \frac{1}{2 \sqrt{3}} \left( \big| u(ds+sd) \big> + \big|  d(su+us) \big> - 2 \big| s(du+ud) \big> \right),\nonumber\\
& \ \ \ \ \big| \phi_{MA} \big> = \frac{1}{2} \left( \big| u(ds-sd) \big> - \big|  d(su-us) \big> \right),\\
&\big|\Sigma^- \big> =  \frac{1}{\sqrt{2}}\left( \big| \phi_{MS} \big> \big| \chi_{MS} \big> +\big|  \phi_{MA} \big> \big| \chi_{MA} \big> \right),\nonumber\\
& \ \ \ \ \big| \phi_{MS} \big> = \frac{1}{\sqrt{6}} \left( \big| d(ds+sd) \big> -2 \big| s dd \big> \right),\nonumber\\
& \ \ \ \ \big| \phi_{MA} \big> = \frac{1}{\sqrt 2} \big|d (ds-sd) \big>,\\
&\big|\Omega_c^0 \big> =  \big| css \big> \big| \chi_{MS}\big>,\\
&\big|\Xi^0 \big> =  \frac{1}{\sqrt{2}}\left( \big| \phi_{MS} \big> \big| \chi_{MS} \big> +\big|  \phi_{MA} \big> \big| \chi_{MA} \big> \right),\nonumber\\
& \ \ \ \ \big| \phi_{MS} \big> = \frac{1}{\sqrt{6}} \left( \big| s(us+su) \big> -2 \big| uss \big> \right),\nonumber\\
& \ \ \ \ \big| \phi_{MA} \big> = -\frac{1}{\sqrt 2} \big|s (us-su)\big>,\\
&\big|\Xi^- \big> =  \frac{1}{\sqrt{2}}\left( \big| \phi_{MS} \big> \big| \chi_{MS} \big> +\big|  \phi_{MA} \big> \big| \chi_{MA} \big> \right),\nonumber\\
& \ \ \ \ \big| \phi_{MS} \big> = -\frac{1}{\sqrt{6}} \left( \big| s(ds+sd) \big> -2 \big| dss \big> \right),\nonumber\\
& \ \ \ \ \big| \phi_{MA} \big> = \frac{1}{\sqrt 2} \big|s (ds-sd) \big>.
\end{align}
\end{subequations}
Here, the $\big| \chi_{MS} \big> $ and $\big| \chi_{MA} \big>$ are the mixed-symmetric and mixed-antisymmetric spin states, respectively, which, together with the symmetric, $\big| \chi_{S} \big> $, and antisymmetric, $\big| \chi_{A} \big> $, states, form an orthogonal basis, such that
\begin{equation}
\big< \chi_{i} \big| \chi_{j} \big> = \delta_{ij}.
\end{equation}

Now we need to go from the charge basis to the isospin basis. When calculating the isospin states, one needs to pay attention to the phases that come from the following isospin multiplets:
\par
\begin{minipage}{0.40\linewidth}
\begin{equation}
\Xi = \begin{pmatrix}
\Xi^0 \\
-\Xi^-
\end{pmatrix},
\end{equation}
 \begin{equation}
\bar{K} = \begin{pmatrix}
\bar{K}^0 \\
-K^-
\end{pmatrix} ,
\end{equation}
\begin{equation}
D = \begin{pmatrix}
D^+ \\
-D^0
\end{pmatrix},
\end{equation}
\end{minipage}
\begin{minipage}{0.45\linewidth}
\begin{equation}
\pi = \begin{pmatrix}
-\pi^+ \\
\pi^0 \\
\pi^-
\end{pmatrix} ,
\end{equation}
\begin{equation}
\Sigma = \begin{pmatrix}
-\Sigma^+ \\
\Sigma^0 \\
\Sigma^-
\end{pmatrix},
\end{equation}
\begin{equation}
\rho = \begin{pmatrix}
-\rho^+ \\
\rho^0 \\
\rho^-
\end{pmatrix}.
\end{equation}
\end{minipage}

Then, for the isospin states we obtain:

\begin{enumerate}
\begin{minipage}{0.57\linewidth}
\item $\big|\Xi_c \pi \big> = \sqrt{\frac{2}{3}}\big|\Xi_c^0 \pi^+ \big> +\sqrt{\frac{1}{3}}\big|\Xi_c^+ \pi^0 \big> ,$
\item $\big|{\Xi'}_c \pi \big> = \sqrt{\frac{2}{3}}\big|{\Xi'}_c^0 \pi^+ \big> +\sqrt{\frac{1}{3}}\big|{\Xi'}_c^+ \pi^0 \big> ,$
\item $\big|\Lambda_c \bar{K} \big> = \big|\Lambda_c^+ \bar{K}^0 \big>, $
\item $\big|\Sigma_c \bar{K} \big> = -\left(\sqrt{\frac{2}{3}}\big|\Sigma_c^{++} K^- \big> + \sqrt{\frac{1}{3}}\big|\Sigma_c^+ \bar{K}^0 \big>\right), $
\item $\big|\Lambda D \big> = \big|\Lambda^0 D^+ \big>, $
\end{minipage}
\begin{minipage}{0.5\linewidth}
\item $\big|\Xi_c \eta \big> = \big|\Xi_c^+ \eta \big> , $
\item $\big|\Sigma D \big> = \sqrt{\frac{2}{3}}\big|\Sigma^+ D^0 \big> -\sqrt{\frac{1}{3}}\big|\Sigma^0 D^+ \big> ,$
\item $\big|{\Xi'}_c \eta \big> = \big|{\Xi'}_c^+ \eta \big> , $
\item $\big|\Omega_c \eta \big> = \big|\Omega_c^0 K^+ \big> , $
\item $\big|\Xi D_s\big> = \big|\Xi^0 D_s^+ \big>.$
\end{minipage}
\end{enumerate}

For the $VB$ sector, the upper vertex of the three vector meson interaction is given by~\cite{Oset:2009vf}
\begin{equation}
\label{eq:VVV}
\mathcal{L}_{VVV} = i g \left< \left( \partial_{\mu} V_{\mu} - \partial_{\nu} V_{\mu}\right) V^{\mu} V^{\nu} \right>,
\end{equation}
and for the lower vertex we again use Eq.~\eqref{eq:VBBql}.

Finally, for the $PB^*$ sector, for the upper vertex we will use again the $VPP$ interaction given by Eq.~\eqref{eq:vpp}. Since from Eq.~\eqref{eq:VBB} to Eq.~\eqref{eq:VBBql} we have made the approximation that $\gamma^{\mu} \rightarrow \gamma^{0}$, this makes Eq.~\eqref{eq:VBBql} spin independent and as such, we can still use it for the $V B^*B^*$ vertices.
Additionally, we have, for the $B^*$ baryons, the following spin-flavor states:
\begin{enumerate}
\begin{minipage}{0.50\linewidth}
\item $\big|\Xi^{*+}_c \big> = \big| \frac{1}{\sqrt{2}} c(us+su) \big> \big| \chi_{S}\big>, $
\item $\big|\Xi^{*0}_c \big> = \big| \frac{1}{\sqrt{2}} c(ds+sd) \big> \big| \chi_{S}\big>, $
\item $\big|\Omega^*_c \big> = \big| css \big> \big| \chi_{S}\big>,$
\item $\big|\Sigma_c^{* ++} \big> =  \big|cuu \big> \big| \chi_{S}\big>, $
\item $\big|\Sigma_c^{* +} \big> =  \big| \frac{1}{\sqrt{2}} c(ud+du) \big> \big| \chi_{S}\big>,$
\end{minipage}
\begin{minipage}{0.5\linewidth}
\item $\big|\Sigma_c^{* 0} \big> =  \big|cdd \big> \big| \chi_{S}\big>, $
\item $\big|\Sigma^{* +} \big> =  \frac{1}{\sqrt{3}}\big|u(su+us) + suu \big> \big| \chi_{S}\big>, $
\item $\big|\Sigma^{* 0} \big> =   \frac{1}{\sqrt{6}} \big|s(du+ud) + d(su+us) +u(sd+ds) \big> \big| \chi_{S}\big>, $
\item $\big|\Sigma^{* -} \big> =  \frac{1}{\sqrt{3}} \big|d(sd+ds) + sdd \big> \big| \chi_{S}\big>.$
\end{minipage}
\end{enumerate}

The isospin states for the $VB$ and $PB^*$ cases are similar to the ones of the $PB$ case.

All these Lagrangians will give similar kernels~\cite{Debastiani:2017ewu,Garzon:2013pad},
\begin{equation}
\label{eq:kernelnr}
V_{ij} = D_{ij} \frac{1}{4 f_{\pi}^2} (p^0+p'^0),
\end{equation}
apart from the different $D_{ij}$ factors that need to be computed.
Here $p^0$ and $p'^0$ are the energies of the initial and final mesons, respectively, and $f_{\pi} = 93$ MeV.
In the case of the $VB$ interaction we get the same kernel, even though the $VVV$ vertex is described by a different Lagrangian, assuming the three momentum of the vectors are small compared to their masses \cite{Oset:2009vf}. Actually the meson baryon chiral lagrangians~\cite{Ecker:1994gg,Bernard:1995dp} can be obtained from the local hidden gauge approach neglecting the $\left(\displaystyle\frac{p}{m_V}\right)^2$ term in the exchanged vectors \cite{Oset:2009vf}. Then, the kernel will be the same as in Eq.~\eqref{eq:kernelnr} with an extra $\vec{\epsilon}\cdot \vec{\epsilon}\ '$ factor, due to the polarizations of the initial and final vector mesons, which can be factorized in the Bethe-Salpeter equation. This means that the equation is spin independent, and that is why we find degenerate states with $J^P=1/2^-$ and $J^P=3/2^-$ with this interaction \cite{Xiao:2013yca}. Because of this, we can just omit that factor.

One can also add relativistic corrections to Eq.~\eqref{eq:kernelnr}, thus obtaining~\cite{Oset2002b},
\begin{equation}
\label{eq:kernelr}
V_{ij} = D_{ij} \frac{2\sqrt{s}-M_{B_i} - M_{B_j}}{4 f_{\pi}^2} \sqrt{\frac{M_{B_i}+E_{B_i}}{2 M_{B_i}}} \sqrt{\frac{M_{B_i}+E_{B_f}}{2 M_{B_f}}},
\end{equation}
where $E_{B_i}, \ E_{B_j}$ are the initial and final baryon energies.

Finally, the $D_{ij}$ coefficients are calculated using the interactions in Eqs.~\eqref{eq:vpp} and~\eqref{eq:VBBql}, and the obtained results are illustrated in Tables~\ref{eq:DijPB}, \ref{eq:DijVB} and \ref{eq:DijPBs} for $PB$, $VB$ and $PB^*$ sectors respectively.
\begin{table*}[h!]
\renewcommand\arraystretch{1.2}
\centering
\caption{\vadjust{\vspace{-2pt}}$D_{ij}$ coefficients for the $PB$ states coupling to $J^P=1/2^-$.}\label{eq:DijPB}
\begin{tabular*}{1.00\textwidth}{@{\extracolsep{\fill}}c|ccccccccccc}
\hline
\hline
$J_{baryon}=1/2$ & $\Xi_c\pi$ & $\Xi_c^\prime\pi$ & $\Lambda_c\bar K$ & $\Sigma_c\bar K$ & $\Lambda D$ & $\Xi_c\eta$ & $\Sigma D$ & $\Xi_c^\prime\eta$ & $\Omega_c K$ & $\Xi D_s$ \\
\hline
$\Xi_c\pi$ & $-2$ & 0 & $-\sqrt\frac{3}{2}$ & 0 & $\frac{1}{2\sqrt 2}\lambda$ & 0 & $-\frac{1}{2\sqrt 2}\lambda$ & 0 & 0 & 0 \\
\hline
$\Xi_c^\prime\pi$ &  & $-2$ & 0 & $-\frac{1}{\sqrt 2}$ & $-\frac{3}{2\sqrt 6}\lambda$ & 0 & $-\frac{1}{2\sqrt 6}\lambda$ & 0 & $-\sqrt 3$ & 0 \\
\hline
$\Lambda_c\bar K$ &  &  & $-1$ & 0 & $-\frac{1}{\sqrt 3}\lambda$ & $\frac{2}{\sqrt 3}$ & 0 & 0 & 0 & 0 \\
\hline
 $\Sigma_c\bar K$ &  &  &  & $-3$ & 0 & 0 & $-\frac{1}{\sqrt 3}\lambda$ & $-2$ & 0 & 0 \\
\hline
$\Lambda D$ &  &  &  &  & $-1$ & $-\frac{1}{6}\lambda$ & 0 & $\frac{1}{2\sqrt 3}\lambda$ & 0 & $-\frac{\sqrt 6}{2}$ \\
\hline
 $\Xi_c\eta$ &  &  &  &  &  & 0 & $-\frac{1}{2}\lambda$ & 0 & 0 & $\frac{1}{\sqrt 6}\lambda$ \\
\hline
 $\Sigma D$ &  &  &  &  &  &  & $-3$ & $-\frac{1}{2\sqrt 3}\lambda$ & 0 & $-\sqrt\frac{3}{2}$ \\
\hline
 $\Xi_c^\prime\eta$ &  &  &  &  &  &  &  & 0 & $-\frac{2\sqrt 6}{3}$ & $-\frac{1}{3\sqrt 2}\lambda$ \\
\hline
 $\Omega_c K$ &  &  &  &  &  &  &  &  & $-2$ & $-\frac{1}{\sqrt 3}\lambda$ \\
\hline
 $\Xi D_s$ &  &  & &  &  &  &  &  &  & $-2$ \\
\hline
\hline
\end{tabular*}
\end{table*}
\begin{table*}[h!]
\renewcommand\arraystretch{1.2}
\centering
\caption{\vadjust{\vspace{-2pt}}$D_{ij}$ coefficients for the $VB$ states coupling to $J^P=1/2^-$, $3/2^-$.}\label{eq:DijVB}
\begin{tabular*}{1.00\textwidth}{@{\extracolsep{\fill}}c|ccccccc}
\hline
\hline
$J_{baryon}=1/2$ & $\Lambda D^*$ & $\Lambda_c\bar K^*$ & $\Sigma D^*$ & $\Xi_c\rho$ & $\Xi_c\omega$ & $\Sigma_c\bar K^*$ \\
\hline
 $\Lambda D^*$ & $-1$ & $-\frac{1}{\sqrt 3}\lambda$ & 0 & $\frac{1}{2\sqrt 2}\lambda$ & $-\frac{1}{2\sqrt 6}\lambda$ & 0 \\
\hline
 $\Lambda_c\bar K^*$ &  & $-1$ & 0 & $-\sqrt\frac{3}{2}$ & $\frac{1}{\sqrt 2}$ & 0 \\
\hline
 $\Sigma D^*$ &  &  & $-3$ & $-\frac{1}{2\sqrt 2}\lambda$ & $-\frac{3}{2\sqrt 6}\lambda$ & $-\frac{1}{\sqrt 3}\lambda$ \\
\hline
 $\Xi_c\rho$ &  &  &  & $-2$ & 0 & 0 \\
\hline
 $\Xi_c\omega$ &  &  &  &  & 0 & 0 \\
\hline
$\Sigma_c\bar K^*$ &  &  &  &  &  & $-3$ \\
\hline
\hline
\end{tabular*}
\end{table*}
\begin{table*}[h!]
\renewcommand\arraystretch{1.2}
\centering
\caption{\vadjust{\vspace{-2pt}}$D_{ij}$ coefficients for the $PB^*$ states coupling to $J^P=3/2^-$.}\label{eq:DijPBs}
\begin{tabular*}{1.00\textwidth}{@{\extracolsep{\fill}}c|cccccc}
\hline
\hline
 $J_{baryon}=3/2$ & $\Xi_c^*\pi$ & $\Sigma_c^*\bar K$ & $\Xi_c^*\eta$ & $\Sigma^*D$ & $\Omega_c^*K$ \\
\hline
 $\Xi_c^*\pi$ & $-2$ & $-\frac{1}{\sqrt 2}$ & 0 & $-\frac{1}{\sqrt 6}\lambda$ & $-\sqrt 3$ \\
\hline
$\Sigma_c^*\bar K$ &  & $-3$ & $-2$ & $\frac{1}{\sqrt 3}\lambda$ & 0 \\
\hline
 $\Xi_c^*\eta$ &  &  & 0 & $-\frac{1}{\sqrt 3}\lambda$ & $-\frac{2\sqrt 6}{3}$ \\
\hline
 $\Sigma^*D$ &  &  &  & $-3$ & 0 \\
\hline
 $\Omega_c^*K$ &  &  & &  & $-2$ \\
\hline
\hline
\end{tabular*}
\end{table*}

\begin{figure}[h!]
  \centering
  \includegraphics[width=0.3\textwidth]{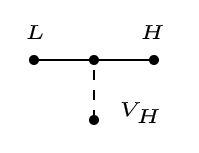}
  \caption{Diagram of $L \rightarrow H$ reaction with $V_H$ emission.}
  \label{fig:feyn2}
\end{figure}

Note the factor $\lambda$ in some of the nondiagonal terms. This factor was added to the terms in the interaction that have an exchange of a heavy vector meson. One can understand this by looking at the propagator of the vector meson ($V_H$) when, in Fig.~\ref{fig:feyn1}, the upper vertex is of the type $L\,  V_H \, H$, with $H, \ V_H$ heavy mesons, and $L$ a light meson (see Fig.~\ref{fig:feyn2}). Then, the propagator will be,
\begin{equation}
\Delta \sim \frac{1}{q^2-m_{V_H}^2},
\end{equation}
with $q$ the transferred momentum. Since $V_H$ is heavy, we can take $\vec{q}\simeq 0$ and
\begin{equation}
\Delta \sim \frac{1}{{(q^0)}^2-m_{V_H}^2} \simeq \frac{1}{(m_H-m_L)^2-m_{V_H}^2},
\end{equation}
where we also have used that $\vec{p}_H \simeq 0$ and $\vec{p}_L \simeq \vec{q} \simeq 0$.
In the calculation of the amplitudes in Eq.~\eqref{eq:kernelnr}, we end up with a factor, relative to the exchange of light vectors
\begin{equation}
\lambda =-m_V^2 \Delta \simeq \frac{-m_V^2}{(m_H-m_L)^2-m_{V_H}^2},
\end{equation}
with $m_V=800\,\rm MeV$. In the case of heavy vector exchange, because $m_{V_H}>m_V$, $\lambda$ will be small and of the order of $\lambda \simeq 1/4$~\cite{Debastiani:2017ewu,Mizutani:2006vq}. We can check that, for the case of $\bar{K} \rightarrow D$ with $D_s^*$ exchange, one gets $\lambda \simeq 0.25$. For more details on how to calculate the kernel of the interaction we refer the reader to the Appendix of Ref.~\cite{Dias:2018qhp}.

Since the $\Xi_c$ and $\Xi_b$ states are heavy quark states, one should comment on how our model deals with HQSS. For that, one should note that, with the exception of the vertices with the $\lambda$, in all other vertices the heavy quark behaves as a spectator, which guarantees that the dominant terms (in the $1/m_Q$ counting) obey HQSS rules. In the terms where that does not happen, their influence is scaled down because of the introduction of the $\lambda$ parameter, which is a small number.

Finally, the same process can be repeated for the beauty sector, where one only needs to substitute the $c$ quark by a $b$ quark. Then, the $D_{ij}$ coefficients will be equal to the ones in the charm sector case, the only difference being that now $\lambda=0.1$ ~\cite{Liang:2017ejq}, because the heavy vector mesons will now be $B^*$ and $B_s^*$ instead of $D^*$ and $D_s^*$.
\section{Results}
With the potential of Eq.~\eqref{eq:kernelr}, the scattering matrix is calculated using the on-shell factorized Bethe-Salpeter equation in coupled channels \cite{Oller:1998zr,Oller:2000fj}
\begin{equation}\label{eq:BS}
T=[1-VG]^{-1}V
\end{equation}
where $V$ is the kernel matrix, $G$ is a diagonal matrix where the diagonal terms correspond to the loop functions of each channel, and the cutoff scheme is used here to regularize the loop integration. The cutoff regularization avoids potential pathologies of the dimensional regularization in the charm sector or beauty sector, where the real part of $G$ can become positive below the threshold and artificial poles can be found in the $T$-matrix, which can lead to the production of the bound states with a repulsive potential \cite{Wu:2010rv}. Also, in order to respect the rules of the heavy quark symmetry in bound states, the same cutoff has to be taken for all channels as it was shown in Refs.~\cite{Ozpineci:2013qza,Lu:2014ina,Altenbuchinger:2013gaa}.

The explicit form for the loop function $G$ is given by \cite{Oset:1997it}
\begin{align}\label{loopfunc}
  G_l=&i\int\frac{d^4q}{(2\pi)^4}\frac{M_l}{E_l(\textbf{q})}\frac{1}{k^0+p^0-q^0-E_l(\textbf{q})+i\epsilon}\frac{1}{q^2-m_l^2+i\epsilon}\nonumber\\
     =&\int\frac{d^3q}{(2\pi)^3}\frac{1}{2\omega_l(\textbf{q})}\frac{M_l}{E_l(\textbf{q})}\frac{1}{k^0+p^0-\omega_l(\textbf{q})-E_l(\textbf{q})+i\epsilon},
\end{align}
which depends on $k^0+p^0=\sqrt s$ and $q_{max}$, and $\omega_l$, $E_l$ are the energies of the meson and baryon respectively and $m_l$, $M_l$ the meson and baryon masses.

Molecular states can be associated with the poles in the scattering amplitudes of Eq.~\eqref{eq:BS} in the complex plane of $\sqrt s$. The poles appearing below the threshold in the first Riemann sheet are categorized as bound states, and those found in the second Riemann sheet are considered to be resonances. The loop function, $G_l$, of a given channel $l$, will be calculated in the first Riemann sheet for Re($\sqrt{s}$) smaller than the threshold of that channel ($\sqrt{s}_{th, l}$), and in the second Riemann sheet for Re($\sqrt{s}$) bigger than $\sqrt{s}_{th, l}$. To take this into account, we define a new loop function
\begin{equation}\label{second loop}
G_l^{\uppercase\expandafter{\romannumeral2}}= \left\{\begin{matrix}
G_l(s) & \text{for } \text{Re}(\sqrt{s})<\sqrt{s}_{th, l}\\
G_l(s)+i\displaystyle\frac{2M_lq}{4\pi\sqrt s} & \text{for } \text{Re}(\sqrt{s})\geq  \sqrt{s}_{th, l}
\end{matrix}\right.,
\end{equation}
where $q$ is given by
\begin{equation}\label{momentum}
q=\frac{\lambda^{1/2}(s,m_l^2,M_l^2)}{2\sqrt s} \quad \text{with Im(q)$>$0},
\end{equation}
with $\lambda(x,y,z)$ the ordinary K\"all\'en function.

After all the poles have been calculated, we can evaluate their coupling constants to various channels. In the vicinity of the poles, the $T$-matrix can be expressed as
\begin{equation}\label{coupling1}
T_{ab}(s)=\frac{g_ag_b}{\sqrt s-z_R},
\end{equation}
where $z_R=M_R-i\Gamma_R/2$ \cite{Garzon:2012np}, which stands for the position of the bound states or resonances found in the complex plane of $\sqrt{s}$. Therefore, the couplings can be evaluated as the residues at the pole of $T_{ab}$, which can be written explicitly with the formula
\begin{equation}\label{coupling2}
g_a^2=\frac{r}{2\pi}\int_0^{2\pi}T_{aa}(z_R+re^{i\theta})e^{i\theta}d\theta.
\end{equation}
However, to be consistent with the different complex phases of each coupling, we choose to calculate the biggest coupling (call it $j$) for each resonance as in Eq.~\eqref{coupling2}, and then calculate the remaining couplings in relation to this one:
\begin{equation}\label{couplings3}
g_a=g_j \lim_{\sqrt{s}\rightarrow z_R} \frac{T_{ja}(s)}{T_{jj}(s)}.
\end{equation}

Meanwhile, with the couplings obtained, we can also evaluate $g_iG^{\uppercase\expandafter{\romannumeral2}}_i$ which can give us the strength of the wave function of the $i$-channel at the origin \cite{Gamermann:2009uq}.

\subsection{Molecular $\Xi_c$ states generated from meson-baryon states}

First, we will start with the $PB$ states, which will lead us to the states with $J^P=1/2^-$. The poles that appear in this sector are illustrated in Table~\ref{pb12c}, where we vary the value of the cutoff $q_{max}$ from $600\,\rm MeV$ to $800\,\rm MeV$.
\begin{table*}[h!]
\renewcommand\arraystretch{1.2}
\centering
\caption{\vadjust{\vspace{-2pt}}Poles in the $J^P=1/2^-$ sector from pseudoscalar-baryon interaction (all units are in MeV).}\label{pb12c}
\begin{tabular*}{1.00\textwidth}{@{\extracolsep{\fill}}ccccccc}
\hline
\hline
 $q_{max}$    &600            &650            &700            &750            &800 \\
\hline
              &$2684.23+i89.72$ &$2679.71+i76.48$ &$2673.49+i64.54$ &$2666.24+i54.01$ &$2658.68+i44.52$ \\
\hline
              &$2800.72+i100.03$&$2801.80+i86.16$ &$2803.28+i72.06$ &$2803.31+i57.77$ &$2794.76+i31.06$ \\
\hline
              &$2880.76+i10.31$ &$2842.47+i10.13$ &$\bm{2791.30+i3.63}$  &$2738.46+i1.36$  &$2685.56+i0.89$ \\
\hline
              &$2896.57+i1.34$  &$2870.10+i10.64$ &$2850.70+i16.38$ &$2830.84+i23.17$ &$2817.77+i40.45$ \\
\hline
              &$2969.50+i3.30$  &$2955.62+i5.10$  &$\bm{2937.15+i7.31}$  &$2913.82+i10.03$ &$2886.31+i13.46$ \\
\hline
              &$3171.55+i32.48$ &$3160.12+i37.77$ &$3148.11+i41.88$ &$3135.67+i44.96$ &$3125.96+i47.18$ \\
\hline
\hline
\end{tabular*}
\end{table*}

It can be seen in Table~\ref{pb12c} that we can always obtain six poles in the range of the masses observed experimentally, and the reason we vary the value of $q_{max}$ is to adjust the pole position to the experimental data. In this way, it can be seen clearly that if we take $q_{max}$ to be $700\,\rm MeV$, we get two poles that are in a good agreement with the experimental data. One is located at $2791.30\,\rm MeV$, the other at $2937.15\,\rm MeV$. They are found to agree very well with the first  and the third resonances of $\Xi_c$, $\Xi_c(2790)$ and $\Xi_c(2930)$, which were first reported in Refs.~\cite{Yelton:2016fqw,Csorna:2000hw} and \cite{Aubert:2007eb} respectively. Although some of the poles in Table~\ref{pb12c} are relatively sensitive to the variation of $q_{max}$, obtaining a good agreement with two resonances simultaneously, while adjusting only one parameter, is very reasonable.  Furthermore, if we look at the imaginary parts of these two poles we get their widths, $7.26\,\rm MeV$ and $14.62\,\rm MeV$, which are very close to the experimental data, $8.9\pm0.6\pm0.8\,\rm MeV$ and $36\pm7\pm11\,\rm MeV$ respectively, within errors. On top of that, it is also important to look at the couplings to various channels as well as the product $g_iG^{\uppercase\expandafter{\romannumeral2}}_i$. As shown in Tables~\ref{pb12cc}, the $2791.30\,\rm MeV$ resonance has a large contribution from the $\Sigma D$ component. However, there are only three open channels where this resonance can decay into, and we can see that one of these open channels $(\Xi_c^\prime\pi)$ is exactly the same channel where the state $\Xi_c(2790)$ was discovered in the first place \cite{Csorna:2000hw,Alexander:1999ud}. Apart from that, although the couplings are considerably smaller than some to closed channels (for example, $\Sigma D$), the coupling constant to $\Xi_c^\prime\pi$ is the dominant one among all the open channels, which is consistent with the experimental observation. Predictions on the decay widths to the open channels can be made precisely using the couplings obtained in Table~\ref{pb12cc} and the formula of Ref.~\cite{Bayar:2017svj}, which has the form
\begin{equation}\label{width}
\Gamma_a=\frac{1}{2\pi}\frac{M_a}{M_R}g_a^2p_a,
\end{equation}
with
\begin{equation}\label{momentum2}
p_a=\frac{\lambda^{1/2}(M_R^2,M_a^2,m_a^2)}{2M_R},
\end{equation}
where $M_a$ and $m_a$ stand for the masses of the $a$th-channel baryon and meson respectively, and $M_R$ is the mass of the resonance (the real part of the pole). In Table~\ref{pole1}, we give the partial decay widths of the poles in Table~\ref{pb12cc},
\begin{table*}[h!]
\renewcommand\arraystretch{1.2}
\centering
\caption{\vadjust{\vspace{-5pt}}The coupling constants to various channels and $g_iG^{\uppercase\expandafter{\romannumeral2}}_i$ for the poles in the $J^P=1/2^-$ sector with $q_{max}=700\,\rm MeV$ (all units are in MeV).}\label{pb12cc}
\begin{tabular*}{\textwidth}{@{\extracolsep{\fill}}cccccc}
\hline
\hline
$\bm{2791.30+i3.63}$& $\Xi_c\pi$ & $\Xi_c^\prime\pi$ & $\Lambda_c\bar K$ & $\Sigma_c\bar K$ & $\Lambda D$ \\
\hline
 $g_i$                &$-0.01-i0.03$&$0.39-i0.44$&$-0.09-i0.05$&$1.05-i0.47$&$1.91-i0.09$\\
 $g_iG^{\uppercase\expandafter{\romannumeral2}}_i$&$0.78+i0.53$&$-3.98+i14.85$&$2.70+i0.73$&$-11.27+i4.95$&$-7.45+i0.27$\\
\hline
                      & $\Xi_c\eta$ & $\Sigma D$ & $\Xi_c^\prime\eta$ & $\Omega_c K$ & $\Xi D_s$\\
\hline
 $g_i$                &$0.23+i0.03$&$\bm{8.82+i0.38}$&$0.49-i0.17$&$0.21-i0.26$&$5.44+i0.20$ \\
 $g_iG^{\uppercase\expandafter{\romannumeral2}}_i$&$-2.00-i0.26$&$\bm{-29.16-i1.48}$&$-3.53+i1.19$&$-1.42+i1.74$&$-11.96-i0.49$\\
 \hline
 \hline
\end{tabular*}
\end{table*}
\begin{table}[h!]
\renewcommand\arraystretch{1.2}
\centering
\caption{\vadjust{\vspace{-2pt}}The widths of pole $2791.30+i3.63$ decaying to various channels (all units are in MeV).}\label{pole1}
\begin{tabular*}{\textwidth}{@{\extracolsep{\fill}}cccc}
\hline
\hline
Channel& $\Xi_c\pi$ & $\Xi_c^\prime\pi$ & $\Lambda_c\bar K$ \\
\hline
 $\Gamma_i$                &0.04&8.00&0.12\\
\hline
\hline
\end{tabular*}
\end{table}
and it can be clearly seen that the state decays mostly to $\Xi_c^\prime\pi$, as expected.

Similarly, for the state located at $2937.15\,\rm MeV$, we can see that the resonance has a large contribution from the $\Lambda D$ channel. Also, we have the same open channels as the ones in Table~\ref{pole1}. We can see that the coupling constant to the channel $\Xi_c^\prime\pi$ becomes smaller than before as shown in Table~\ref{pb12ccc}. However, the couplings to the channels $\Xi_c\pi$ and $\Lambda_c\bar K$ are bigger, yet, there is more phase space for decay for $\Xi_c\pi$ and $\Xi_c^\prime\pi$, but altogether the final widths to these three channels are comparable as one can see in Table~\ref{pole2}. The $\Lambda_c\bar K$ channel accounts for about $1/3$ of the total width and this is the channel where the BaBar Collaboration observed the state $\Xi_c(2930)$ \cite{Aubert:2007eb}.
\begin{table*}[h!]
\renewcommand\arraystretch{1.2}
\centering
\caption{\vadjust{\vspace{-2pt}}The coupling constants to various pseudoscalar-baryon channels and $g_iG^{\uppercase\expandafter{\romannumeral2}}_i$ for the poles in the $J^P=1/2^-$ sector with $q_{max}=700\,\rm MeV$ (all units are in MeV).}\label{pb12ccc}
\begin{tabular*}{\textwidth}{@{\extracolsep{\fill}}cccccc}
\hline
\hline
$\bm{2937.15+i7.31}$& $\Xi_c\pi$ & $\Xi_c^\prime\pi$ & $\Lambda_c\bar K$ & $\Sigma_c\bar K$ & $\Lambda D$ \\
\hline
 $g_i$                &$-0.29+i0.10$&$0.03-i0.32$&$0.28-i0.22$&$0.27+i0.08$&$\bm{3.96-i0.29}$\\
 $g_iG^{\uppercase\expandafter{\romannumeral2}}_i$&$0.83-i9.44$&$6.42+i6.63$&$0.31+i10.35$&$-5.40-i1.98$&$\bm{-27.75+i0.73}$\\
\hline
                      & $\Xi_c\eta$ & $\Sigma D$ & $\Xi_c^\prime\eta$ & $\Omega_c K$ & $\Xi D_s$\\
\hline
 $g_i$                &$-0.07+i0.39$&$-2.44+i0.10$&$0.09+i0.04$&$-0.12-i0.26$&$3.55-i0.13$ \\
 $g_iG^{\uppercase\expandafter{\romannumeral2}}_i$&$1.11-i5.19$&$12.15-i0.15$&$-0.85-i0.40$&$0.98+i2.23$&$-9.83+i0.23$\\
 \hline
 \hline
\end{tabular*}
\end{table*}
\begin{table}[h!]
\renewcommand\arraystretch{1.2}
\centering
\caption{\vadjust{\vspace{-2pt}}The widths of pole $2937.15+i7.31$ decaying to various channels (all units are in MeV).}\label{pole2}
\begin{tabular*}{\textwidth}{@{\extracolsep{\fill}}cccc}
\hline
\hline
Channel& $\Xi_c\pi$ & $\Xi_c^\prime\pi$ & $\Lambda_c\bar K$ \\
\hline
 $\Gamma_i$                &5.22&4.45&5.88\\
 \hline
\hline
\end{tabular*}
\end{table}

On the other hand, for the $VB$ channels, in Table~\ref{vbp}, we obtain four poles for all the cutoffs, and in order to be consistent with the $J^P=1/2^-$ sector, we stick to the same cutoff $q_{max}=700\,\rm MeV$, which leads us to three poles that can be selected as possible candidates for $\Xi_c(2970)$, $\Xi_c(3055)$ or $\Xi_c(3080)$, and $\Xi_c(3123)$ states.
\begin{table*}[h!]
\renewcommand\arraystretch{1.2}
\centering
\caption{\vadjust{\vspace{-2pt}}The poles in the $J^P=1/2^-$, $3/2^-$ sector from the vector-baron interaction (all units are in MeV).}\label{vbp}
\begin{tabular*}{1.00\textwidth}{@{\extracolsep{\fill}}ccccccc}
\hline
\hline
 $q_{max}$    &600            &650            &700            &750            &800 \\
\hline
              &3055.63       &3016.46 &$\bm{2973.76}$ &2928.28 &2880.75 \\
\hline
              &3117.37       &3094.39 &$\bm{3068.21}$ &3040.89 &3013.14 \\
\hline
              &3121.75       &3115.67 &$\bm{3109.04}$ &3100.55 &3090.16 \\
\hline
              &3234.03+i0.22 &3204.98 &3174.50 &3143.09 &3111.43 \\
\hline
\hline
\end{tabular*}
\end{table*}
\begin{table*}[h!]
\renewcommand\arraystretch{1.2}
\centering
\caption{\vadjust{\vspace{-2pt}}The coupling constants to various vector-baryon channels and $g_iG^{\uppercase\expandafter{\romannumeral2}}_i$ for the poles in the $J^P=1/2^-,3/2^-$ sector with $q_{max}=700\,\rm MeV$ (all units are in MeV).}\label{vbcc}
\begin{tabular*}{\textwidth}{@{\extracolsep{\fill}}ccccccc}
\hline
\hline
$\bm{2973.76}$&$\Lambda D^*$ & $\Lambda_c\bar K^*$ & $\Sigma D^*$ & $\Xi_c\rho$ & $\Xi_c\omega$ & $\Sigma_c\bar K^*$\\
\hline
 $g_i$                &0&0.07&$\bm{9.30}$&0.33&0.30&0.55\\
 $g_iG^{\uppercase\expandafter{\romannumeral2}}_i$&0&$-0.48$&$\bm{-31.85}$&$-2.29$&$-2.02$&$-2.87$\\
\hline
\hline
$\bm{3068.21}$&$\Lambda D^*$ & $\Lambda_c\bar K^*$ & $\Sigma D^*$ & $\Xi_c\rho$ & $\Xi_c\omega$ & $\Sigma_c\bar K^*$\\
\hline
 $g_i$                &0.37&$\bm{3.08}$&$-0.26$&$\bm{3.57}$&$-0.85$&$-0.04$\\
 $g_iG^{\uppercase\expandafter{\romannumeral2}}_i$&$-2.33$&$\bm{-30.22}$&1.20&$\bm{-30.89}$&7.19&0.22\\
\hline
\hline
$\bm{3109.04}$&$\Lambda D^*$ & $\Lambda_c\bar K^*$ & $\Sigma D^*$ & $\Xi_c\rho$ & $\Xi_c\omega$ & $\Sigma_c\bar K^*$\\
\hline
 $g_i$                &$\bm{3.05}$&0.05&0.03&$-0.51$&0.09&0.01\\
 $g_iG^{\uppercase\expandafter{\romannumeral2}}_i$&$\bm{-26.23}$&$-0.60$&$-0.17$&5.04&$-0.81$&$-0.05$\\
\hline
\hline
\end{tabular*}
\end{table*}

As shown in Table~\ref{vbcc}, we present the couplings of the first three poles for $q_{max}=700\,\rm MeV$. The first state, $2973.76\,\rm MeV$ couples very strongly to $\Sigma D^*$ and almost nothing to the rest of the channels, thus it can be considered as a $\Sigma D^*$ bound state. The second state, located at $3068.21\,\rm MeV$, couples to both $\Lambda_c \bar K^*$ and $\Xi_c\rho$, with similar values for the coupling as well as $gG^{\uppercase\expandafter{\romannumeral2}}$. The situation of the third state, $3109.04\,\rm MeV$, is similar to what we found in the first state, where it practically only couples to $\Lambda D^*$, and the product $gG^{\uppercase\expandafter{\romannumeral2}}$ is also significantly larger than for the rest of the channels. Moreover, we notice that all these three poles are below thresholds, so they do not decay to any of the coupled states shown in Table~\ref{vbcc}, instead it may decay into the pseudoscalar-baryon ones.

Now we study the $PB^*$ states with $J^P=3/2^-$. All the poles obtained in this sector are given in Table~\ref{pb32}. One can see from the poles with $q_{max}=700\,\rm MeV$, two of them are relatively close to the experimental observed states, $\Xi_c(2930)$ and $\Xi_c(3055)$. It is noteworthy that the first one, $2912.78+i19.94$, which agrees rather well the experimental data, is similar to the pole $(2937.15+i7.31)$ found in the $J^P=1/2^-$ sector in Table~\ref{pb12c}, since the $J^P$ of $\Xi_c(2930)$ has not been measured, it can be either one of these two poles. Although the mass of the state, $2912.78\,\rm MeV$, may be slightly smaller than the observed state $\Xi_c(2930)$, the width (which is $39.88\,\rm MeV$) is remarkably close to the data fit ($36\pm7\pm11\,\rm MeV$) by the BaBar Collaboration \cite{Aubert:2007eb}.
Similarly, for the other pole obtained at the position $3015.18\,\rm MeV$, its width ($2.74\,\rm MeV$) is a few MeV below the value of $7.8\pm1.2\pm1.5\,\rm MeV$ reported in Refs.~\cite{Tanabashi:2018oca,Kato:2016hca}. Apart from that, in Table~\ref{pb32c}, we present our results for the couplings of these two poles to various channels, where we can see that for the first pole only the channel $\Xi_c^*\pi$ is open for decay, and the width of this state decaying to $\Xi_c^*\pi$ is found to be $56.92\,\rm MeV$ using Eq.~\eqref{width}. On the other hand, for the pole at $3015.18\,\rm MeV$, both channels $\Xi_c^*\pi$ and $\Sigma_c^*\bar K$ are open for decay. We can see that the state at $3015.18\,\rm MeV$ couples mostly to the channel $\Sigma^*D$, and the coupling constants to the open channels are both very small, but the strengths of the wave functions at the origin are considerable for these open channels, and clearly, $gG^{\uppercase\expandafter{\romannumeral2}}$ for the former channel is bigger than the latter one, which can also be seen in the widths obtained, $2.65\,\rm MeV$ and $0.08\,\rm MeV$, for the channels $\Xi_c^*\pi$ and $\Sigma_c^*\bar K$, respectively.
\begin{table*}[h!]
\renewcommand\arraystretch{1.2}
\centering
\caption{\vadjust{\vspace{-2pt}}The poles in the $J^P=3/2^-$ sector from the pseudoscalar-baron interaction (all units are in MeV).}\label{pb32}
\begin{tabular*}{1.00\textwidth}{@{\extracolsep{\fill}}ccccccc}
\hline
\hline
 $q_{max}$    &600             &650            &700             &750            &800 \\
\hline
              &$2868.84+i101.02$ &$2869.69+i87.71$ &$2870.00+i71.15$  &$2871.12+i55.04$ &$2888.93+i43.98$ \\
\hline
              &$2950.39+i11.19$  &$2932.11+i15.01$ &$\bm{2912.78+i19.94}$  &$2891.71+i27.88$ &$2855.31+i26.46$ \\
\hline
              &$3099.36+i0.55$   &$3059.03+i0.89$  &$\bm{3015.18+i1.37}$   &$2968.69+i2.98$  &$2918.23+i7.32$ \\
\hline
              &$3243.94+i32.64$  &$3233.36+i38.32$ &$3222.35+i42.93$  &$3211.05+i46.56$ &$3199.61+i49.32$ \\
\hline
\hline
\end{tabular*}
\end{table*}
\begin{table*}[h!]
\renewcommand\arraystretch{1.2}
\centering
\caption{\vadjust{\vspace{-2pt}}The coupling constants to various pseudoscalar-baryon channels and $g_iG^{\uppercase\expandafter{\romannumeral2}}_i$ for the poles in the $J^P=3/2^-$ sector with $q_{max}=700\,\rm MeV$ (all units are in MeV).}\label{pb32c}
\begin{tabular*}{\textwidth}{@{\extracolsep{\fill}}cccccc}
\hline
\hline
$\bm{2912.78+i19.94}$& $\Xi_c^*\pi$ & $\Sigma_c^*\bar K$ & $\Xi_c^*\eta$ & $\Sigma^*D$ & $\Omega_c^*K$ \\
\hline
 $g_i$                &$0.41-i1.28$&$\bm{3.78-i0.47}$&$1.87-i0.12$&$-1.09-i0.83$&$0.16-i0.85$\\
 $g_iG^{\uppercase\expandafter{\romannumeral2}}_i$&$11.65+i36.25$&$\bm{-48.53+i2.60}$&$-14.80+i0.42$&$3.17+i2.60$&$-1.34+i6.20$\\
\hline
\hline
$\bm{3015.18+i1.37}$& $\Xi_c^*\pi$ & $\Sigma_c^*\bar K$ & $\Xi_c^*\eta$ & $\Sigma^*D$ & $\Omega_c^*K$ \\
\hline
 $g_i$                &$0.03-i0.24$&$0.11+i0.07$&$0.42+i0.03$&$\bm{8.94-i0.04}$&$-0.04-i0.20$\\
 $g_iG^{\uppercase\expandafter{\romannumeral2}}_i$&$4.82+i5.04$&$-3.19-i1.64$&$-4.18-i0.34$&$\bm{-33.36+i0.03}$&$0.31+i1.72$\\
\hline
\hline
\end{tabular*}
\end{table*}
\subsection{Molecular states for $\Xi_b$ generated from meson-baryon states}
In this subsection, we follow closely the calculations in the previous one. It starts with the $J^P=1/2^-$ sector, where we also have ten coupled channels similar to the ones we considered for $\Xi_c$, only with the $c$ quark replaced by a $b$ quark in each channel. The poles from the $PB$ interaction are given in Table~\ref{pb12b}, where we obtain six poles for each cutoff. Taking into account the uncertainty caused by the variation of the cutoff, we can associate the pole, $6220.30\,\rm MeV$ with $q_{max}=650\,\rm MeV$, to the state $\Xi_b(6226)$ recently observed by the LHCb Collaboration \cite{Aaij:2018yqz}. The newly observed state $\Xi_b(6226)$ is reported with the values $6226.9\pm2.0\pm0.3\pm0.2\,\rm MeV/c^2$ and $18.1\pm5.4\pm1.8\,\rm MeV/c^2$ for its mass and width, respectively. We can see that the mass obtained is merely a few MeV below the experimental data, and the width ($25.20\,\rm MeV$) is also in very good agreement with the data.
\begin{table*}[h!]
\renewcommand\arraystretch{1.2}
\centering
\caption{\vadjust{\vspace{-2pt}}The poles in the $J^P=1/2^-$ sector from the pseudoscalar-baron interaction (all units are in MeV).}\label{pb12b}
\begin{tabular*}{1.00\textwidth}{@{\extracolsep{\fill}}ccccccc}
\hline
\hline
 $q_{max}$    &600            &650            &700            &750            &800 \\
\hline
              &$6002.21+i81.90$ &$5997.45+i69.73$ &$5991.25+i58.74$ &$5984.13+i49.02$ &$5976.61+i40.29$ \\
\hline
              &$6152.19+i91.66$ &$6152.17+i78.48$ &$6152.24+i64.42$ &$6150.40+i48.36$ &$6137.15+i27.48$ \\
\hline
              &$6237.52+i11.30$ &$\bm{6220.30+i12.60}$ &$6201.74+i19.00$ &$6183.11+i27.85$ &$6175.03+i40.45$ \\
\hline
              &$6263.48+i0.07$  &$6205.08+i2.94$  &$6141.06+i1.73$  &$6073.96+i0.17$  &$6004.35+i0.44$ \\
\hline
              &$6359.89+i0.82$  &$6338.97+i1.44$  &$6312.50+i2.42$  &$6280.97+i3.77$  &$6244.54+i5.65$ \\
\hline
              &$6513.45+i29.56$ &$6501.26+i33.87$ &$6488.63+i37.17$ &$6482.88+i39.75$ &$6482.86+i41.90$ \\
\hline
\hline
\end{tabular*}
\end{table*}
\begin{table*}[h!]
\renewcommand\arraystretch{1.2}
\centering
\caption{\vadjust{\vspace{-2pt}}The coupling constants to various pseudoscalar-baryon channels and $g_iG^{\uppercase\expandafter{\romannumeral2}}_i$ for the poles in the $J^P=1/2^-$ sector with $q_{max}=650\,\rm MeV$ (all units are in MeV).}\label{pb12bc}
\begin{tabular*}{\textwidth}{@{\extracolsep{\fill}}cccccc}
\hline
\hline
$\bm{6220.30+i12.60}$& $\Xi_b\pi$ & $\Xi_b^\prime\pi$ & $\Lambda_b\bar K$ & $\Sigma_b\bar K$ & $\Lambda \bar B$ \\
\hline
 $g_i$                &$0.01+i0.02$&$0.34-i0.91$&$0.01-i0.01$&$\bm{3.53-i0.14}$&$-1.03+i0.61$\\
 $g_iG^{\uppercase\expandafter{\romannumeral2}}_i$&$-0.60+i0.05$&$10.08+i25.84$&$0.42+i0.15$&$\bm{-44.85-i0.67}$&$1.47-i0.78$\\
\hline
                      & $\Xi_b\eta$ & $\Sigma \bar B$ & $\Xi_b^\prime\eta$ & $\Omega_b K$ & $\Xi B_s$\\
\hline
 $g_i$                &$-0.00+i0.04$&$-2.09+i4.72$&$1.80+i0.02$&$0.09-i0.65$&$-1.93+i2.81$ \\
 $g_iG^{\uppercase\expandafter{\romannumeral2}}_i$&$0.02-i0.38$&$2.54-i5.25$&$-13.14-i0.55$&$-0.74+i4.33$&$1.45-i2.02$\\
\hline
\hline
\end{tabular*}
\end{table*}
Moreover, for the couplings of this pole we can look at the results in Table~\ref{pb12bc}, where we can see that the main contribution comes from the $\Sigma_b \bar K$ channel \footnote{The large attraction in the $\Sigma_b K^-$ channel and the possibility of having a bound state with this channel has been discussed before in Refs.~\cite{Fan:2011aa,Lu:2014ina,Huang:2018bed}}. Also, we see that it has only three open channels, and it should be noted that two of these channels, the $\Xi_b\pi$ and $\Lambda_b\bar K$, are the ones where the state $\Xi_b(6226)$ has been observed \cite{Aaij:2018yqz}. However, according to our findings, it couples mostly to the $\Xi_b^\prime\pi$ among the open channels, which suggests that it would be easier to find the state $\Xi_b(6226)$ in the $\Xi_b^\prime\pi$ channel instead of the other two, which can be confirmed by future experiments.
\begin{table}[h!]
\renewcommand\arraystretch{1.2}
\centering
\caption{\vadjust{\vspace{-2pt}}The widths of pole $6220.30+i12.60$ decaying to various channels (all units are in MeV).}\label{pole3}
\begin{tabular*}{\textwidth}{@{\extracolsep{\fill}}cccc}
\hline
\hline
Channel& $\Xi_b\pi$ & $\Xi_b^\prime\pi$ & $\Lambda_b\bar K$ \\
\hline
 $\Gamma_i$                &0.02&35.01&0.01\\
 \hline
\hline
\end{tabular*}
\end{table}
The decay widths of the pole of Table~\ref{pb12bc} are given in Table~\ref{pole3}, where we can obtain the ratio of branching fractions of state $6220.30\,\rm MeV$ as follows
\begin{equation}
\frac{\mathcal B(6220.30\to\Lambda_b\bar K)}{\mathcal B(6220.30\to\Xi_b\pi)}=0.50,
\end{equation}
which is relatively close to the results ($1\pm0.5$) presented in Ref.~\cite{Aaij:2018yqz} given its uncertainty.

On the other hand, for the $VB$ interaction, we observed four poles for each cutoff, but none of these poles for $q_{max}=650\,\rm MeV$ can be associated to any known $\Xi_b$ states with negative parity, since there are not enough data available for $\Xi_b$ states. Furthermore, almost all of the poles found in this sector are below their respective thresholds, which makes it more plausible that these channels could qualify as bound states rather than resonances.
\begin{table*}[h!]
\renewcommand\arraystretch{1.2}
\centering
\caption{\vadjust{\vspace{-2pt}}The poles in the $J^P=1/2^-$, $3/2^-$ sector from the vector-baron interaction (all units are in MeV).}\label{vbpole}
\begin{tabular*}{1.00\textwidth}{@{\extracolsep{\fill}}ccccccc}
\hline
\hline
 $q_{max}$    &600            &650            &700            &750            &800 \\
\hline
              &6342.09       &6295.86        &6244.94         &6190.01 &6131.80 \\
\hline
              &6425.22       &6407.58        &6379.79         &6351.13 &6322.26 \\
\hline
              &6434.39       &6417.24        &6406.58         &6393.82 &6378.88 \\
\hline
              &$6579.47+i0.05$ &$6548.79+i0.03$  &6516.72         &6484.01 &6451.25 \\
\hline
\hline
\end{tabular*}
\end{table*}

Moving on to the $J^P=3/2^-$ sector, we also get four poles for each cutoff, which are given in Table~\ref{pbpole}, where we also find a possible candidate for $\Xi_b(6227)$. The state $6240.21\,\rm MeV$ agrees really well with the experimental data \cite{Aaij:2018yqz}, as both mass and width are within acceptable ranges. For the couplings as well as $g_iG^{\uppercase\expandafter{\romannumeral2}}_i$, shown in Table~\ref{pb32bc}, it can be seen that the state at $6240.21\,\rm MeV$ couples mostly to $\Sigma_b^*\bar K$ and $\Xi_b^*\eta$, and only slightly to the rest of the channels. However,
\begin{table*}[h!]
\renewcommand\arraystretch{1.2}
\centering
\caption{\vadjust{\vspace{-2pt}}The poles in the $J^P=3/2^-$ sector from the pseudoscalar-baron interaction (all units are in MeV).}\label{pbpole}
\begin{tabular*}{1.00\textwidth}{@{\extracolsep{\fill}}ccccccc}
\hline
\hline
 $q_{max}$    &600             &650            &700             &750            &800 \\
\hline
              &$6169.97+i92.88$  &$6169.85+i80.29$ &$6168.88+i66.55$  &$6166.89+i49.80$ &$6155.48+i29.00$ \\
\hline
              &$6258.53+i10.88$ &$\bm{6240.21+i14.65}$ &$6221.49+i19.71$  &$6203.04+i27.94$ &$6193.81+i39.67$ \\
\hline
              &$6474.16+i0.14$   &$6424.37+i0.20$  &$6369.22+i0.34$   &$6309.51+i0.68$  &$6245.85+i2.08$ \\
\hline
              &$6538.85+i30.15$  &$6527.01+i34.64$ &$6514.86+i38.05$  &$6502.41+i40.62$ &$6500.43+i42.56$ \\
\hline
\hline
\end{tabular*}
\end{table*}
\begin{table*}[h!]
\renewcommand\arraystretch{1.2}
\centering
\caption{\vadjust{\vspace{-2pt}}The coupling constants to various pseudoscalar-baryon channels and $g_iG^{\uppercase\expandafter{\romannumeral2}}_i$ for the poles in the $J^P=3/2^-$ sector with $q_{max}=650\,\rm MeV$ (all units are in MeV).}\label{pb32bc}
\begin{tabular*}{\textwidth}{@{\extracolsep{\fill}}cccccc}
\hline
\hline
$\bm{6240.21+i14.65}$& $\Xi_b^*\pi$ & $\Sigma_b^*\bar K$ & $\Xi_b^*\eta$ & $\Sigma^*\bar B$ & $\Omega_b^*K$ \\
\hline
 $g_i$                &$0.23-i0.93$&$\bm{3.39-i0.36}$&$1.74-i0.10$&$-0.78-i0.36$&$0.03-i0.65$\\
 $g_iG^{\uppercase\expandafter{\romannumeral2}}_i$&$13.03+i24.31$&$\bm{-43.16+i1.85}$&$-12.78+i0.27$&$0.63+i0.30$&$-0.29+i4.27$\\
\hline
\hline
\end{tabular*}
\end{table*}
when we look at the magnitude of $gG^{\uppercase\expandafter{\romannumeral2}}$, we can see that not only the channel $\Sigma_b^*\bar K$ is significantly bigger than the others, also the only open channel $\Xi_b^*\pi$ is considerably large compared to the other channels. Besides, the decay width of this particular pole to $\Xi_b^*\pi$ is $34.3851\,\rm MeV$, which is similar to the value of the width for at $6220.30$ MeV decaying to $\Xi_b^\prime\pi$, in Table~\ref{pole3}, because they have almost the same phase space for decay.

\section{Summary and discussion}

Motivated by the experimental findings of $\Xi_c$ and $\Xi_b$ states, we use the Bethe-Salpeter coupled channel formalism to study the $\Xi_{c(b)}$ states dynamically generated from the meson-baryon interaction, considering three types of interactions ($PB$, $VB$ and $PB^*$), for both the charm and beauty sectors. We search for the pole with different cutoffs in the second Riemann sheet once the scattering matrix is evaluated. Apart from that, the couplings of the poles to various channels are also calculated. With that, we are able the assess the strength at the origin of the wave function and further evaluate the decay widths to the open channels.

The only free parameter in our study is the loop regulator in the meson-baryon loop function, where we employ the cutoff regularization scheme, and we have taken different values for the cutoff in the charm and beauty sectors.

We obtain multiple states for $\Xi_c$, with some of them agreeing significantly well with the experimental data. For example, the lowest state we observe in the charm sector is the state at $2791.30\,\rm MeV$ (with the width $7.26\,\rm MeV$) generated from the pseudoscalar-baryon interaction, which have the same $J^P$ quantum numbers as the state $\Xi_c(2790)$ (with width $8.9\pm0.6\pm0.8\,\rm MeV$). It can also be seen that there is a very good agreement in their masses and widths. On top of that, we also obtain states at $2937.15\,\rm MeV$ ($2912.78\,\rm MeV$), $2973.76\,\rm MeV$, $3068.21\,\rm MeV$ ($3015.18\,\rm MeV$) and $3109.04\,\rm MeV$, which can be associated to the experimentally observed states $\Xi_c(2930)$, $\Xi_c(2970)$, $\Xi_c(3055)$ and $\Xi_c(3080)$, respectively. On the other hand, we found two poles, at $6220.30\,\rm MeV$ (with a width $25.20\,\rm MeV$) and $6240.21\,\rm MeV$ (with width $29.30\,\rm MeV$) in the $1/2^-$ and $3/2^-$ sectors, respectively. We can see that both their masses and widths agree well with the recently observed state $\Xi_b(6227)$ with a width $18.1\pm5.4\pm1.8\,\rm MeV$.

Overall, the states obtained in this work agree well with some of the already observed states in both the charm and beauty sectors, and it would be interesting to see if further measurements of spin and parity of these states would also agree with our predicted states. Furthermore, with the increased luminosity in future runs, the comparisons of the predictions made here and the experimental measurements will shed light on the nature of these hadrons.
\section{ACKNOWLEDGEMENT}
Q.~X.~Yu acknowledges the support from the National Natural Science Foundation of China (Grant No.~11775024 and 11575023). R.~P and V.~R.~D. wish to acknowledge the Generalitat Valenciana in the program Santiago Grisolia. This work is partly supported
by the Spanish Ministerio de Economia y Competitividad and European FEDER funds
under Contracts No. FIS2017-84038-C2-1-P B and No. FIS2017-84038-C2-2-P B, and the
Generalitat Valenciana in the program Prometeo II-2014/068, and the project Severo Ochoa
of IFIC, SEV-2014-0398.


\end{document}